\documentclass[epj]{svjour}
\usepackage{graphics}
\begin{document}
\title{Vacuum polarization in muonic atoms: the Lamb shift at low and medium $Z$}

\author{Savely G. Karshenboim\inst{1,2,}\thanks{E-mail: sek@mpq.mpg.de}
   \and Vladimir G. Ivanov\inst{1,3,}
   \and Evgeny Yu. Korzinin\inst{1,}
}                     
\offprints{}          
\institute{D. I. Mendeleev Institute for Metrology (VNIIM),
190005, St. Petersburg, Russia
 \and Max-Planck-Institut f\"ur Quantenoptik, 85748, Garching, Germany
 \and Pulkovo Observatory, 196140, St. Petersburg, Russia}
\date{Received: date / Revised version: date}
%
\abstract{ In muonic atoms the Uehling potential (an effect of a
free electronic vacuum polarization loop) is responsible for the
leading contribution to the Lamb shift causing the splitting of
states with $\Delta n=0$ and $\Delta l\neq0$. Here we consider the
Lamb shift in the leading nonrelativistic approximation, i.e.,
within an approach based on a certain Schr\"odinger equation. That
is valid for low and medium $Z$ as long as $(Z\alpha)^2\ll1$. The
result is a function of a few parameters, including
$\kappa=Z\alpha m_\mu/m_e$, $n$ and $l$. We present various
asymptotics and in particular we study a region of validity of
asymptotics with large and small $\kappa$. Special attention is
paid to circular states, which are considered in a limit of
$n\gg1$.
} 
\maketitle
\section{Introduction}
\label{intro} The gross structure of energy levels in all kinds of
hydrogen-like atoms is generally of the same form determined by
the Schr\"odinger-Coulomb equation
\[
E (nl_j) \simeq - \frac{(Z\alpha)^2mc^2}{2n^2}\;,
\]
where $m$ is the mass of the orbiting particle which is an
electron in a conventional atom and a heavier particle in a muonic
or exotic atom. However, details of the spectrum and, in
particular, the structure of the energy levels with the same value
of the principal quantum number $n$ are different in different
kinds of atoms. For example, in muonic atoms at low and medium $Z$
the largest splitting between states with the same $n$ is the one
for states with $\Delta l\neq0$ (the Lamb splitting) which is
essentially a nonrelativistic effect.

In the nonrelativistic approximation the leading contribution to
the Lamb shift in muonic atoms (i. e., the Uehling correction) has
been known analytically for a while for certain levels
\cite{Pusto}, however, only numerical results used to be quoted in
the textbooks (see, e.g., \cite{IV}). A reason for that is the
complicated form of the analytic expressions. For instance, in the
simplest case of the ground state the result is of the form
\cite{Pusto}
\begin{eqnarray}\label{eu1s}
\Delta E(1s) &=& - \frac{\alpha}{3\pi}\,(Z\alpha)^2mc^2\,\biggl\{
-\frac{4+\kappa^2-2\,\kappa^4}{\kappa^3}\cdot A(\kappa)
\nonumber\\
 &+&
\frac{4+3\,\kappa^2}{\kappa^3}\cdot \frac{\pi}{2}
-\frac{12+11\,\kappa^2}{3\,\kappa^2} \biggr\}\;,
\end{eqnarray}
where
\begin{equation}
  A(\kappa)=\frac{\arccos(\kappa)}{\sqrt{1-\kappa^2}}=
  \frac{\ln\left(\kappa+\sqrt{\kappa^2-1}\right)}{\sqrt{\kappa^2-1}}\,,\nonumber\\
\end{equation}
\begin{equation}
  \kappa = \frac{Z\alpha \, m}{m_e} \,,\nonumber\\
\end{equation}
and $m_e$ is the electron mass. Expressions for other states are
similar, but more complicated. They involve functions
$A(\kappa_n)$ with a characteristic parameter
\begin{equation}
  \kappa_n = \frac{\kappa}{n}
  \,,\nonumber\\
\end{equation}
and coefficients similar to those in Eq.(\ref{eu1s}) depend on
values of the principal and orbital quantum numbers, $n$ and $l$.

The mass of the orbiting particle $m$ in a non-conventional
hydrogen-like atom is much above the electron mass $m_e$. We
consider here the vacuum polarization effects for a hydrogen-like
atoms with an orbiting particle, which  in particular may be a
muon ($m_\mu\simeq 207\, m_e$; $\kappa\simeq 1.5\, Z$), a pion
($m_\pi\simeq 273\, m_e$; $\kappa\simeq 2\, Z$), an antiproton
(${m_{\overline{p}}\simeq 1836\, m_e}$; $\kappa\simeq 13\,Z$) etc.
The relativistic effects for those atoms are quite different for
various reasons, while the result in the leading nonrelativistic
approximation is the same. Further we do not distinguish between
various possibilities of the orbiting particles and mainly speak
about a muon, but the equations could be applied to any orbiting
particle.

Analytic results have been known for some time even for
hydrogen-like atoms with a Dirac particle \cite{CJP98} and since
recently for the case of a Klein-Gordon particle \cite{cjp_kg}.
They are rather cumbersome, containing the hypergeometric function
$_3F_2$ and far from being transparent. For instance, the
relativistic result \cite{CJP98,cjp_kg} for the $nl$ states reads
as a finite sum over basic integrals
\begin{eqnarray}\label{defKabc}
K_{abc}(\widetilde{\kappa}_n)&=&
  \frac{1}{2}\widetilde{\kappa}_n^c\,
  B\bigl(a+1/2,1-b/2+c/2\bigr)
\nonumber\\
  &\times &
  {_3F_2}\bigl(c/2,\, c/2+1/2,\, 1-b/2+c/2 ;
  \nonumber\\
  &\ &\qquad
  1/2,\, a+3/2-b/2+c/2 ;\; \widetilde{\kappa}_n^2\bigr)
 \\
  &-&  \frac{c}{2}\,\widetilde{\kappa}_n^{c+1}\,
  B\bigl(a+1/2, 3/2-b/2+c/2\bigr)
  \nonumber\\
  &\times &
  {_3F_2}\bigl(c/2+1,\, c/2+1/2,\, 3/2-b/2+c/2;
  \nonumber\\
  &\ &\qquad
  3/2,\, a+2-b/2+c/2;\; \widetilde{\kappa}_n^2\bigr)
\;,\nonumber
\end{eqnarray}
where
${_3F_2}\bigl(\alpha_1,\alpha_2,\alpha_3;\;\beta_1,\beta_2;\;
z\bigr)$ stands for the generalized hypergeometric function (see,
e.g., \cite{3f2} and $B\bigl(\alpha_1,\alpha_2\bigr)$ is the beta
function. The parameters $a,b,c$ are linear functions of $n$ and
$l$ in the nonrelativistic case, while for the relativistic
results they contain certain additions of relativistic corrections
which go to zero at the limit of $(Z\alpha)\to 0$. The argument of
${_3F_2}$, $\widetilde{\kappa}_n^2$, is reduced to $\kappa_n^2$ in
the nonrelativistic approximation.

Meanwhile, muonic atoms offer a special region of parameters where
the result can be essentially simplified (see e.g.
\cite{EJP,CJP98}). For instance, the Uehling correction for the
ground state \cite{EJP,CJP98} (cf. Eq.(\ref{eu1s})) takes the form
\begin{equation} \label{as1s}
  \Delta E(1s) \simeq
  -\frac{\alpha}{\pi}
  \,
  (Z\alpha)^2mc^2
  \,
  \left(
   \frac{2}{3}\ln \bigl(2\kappa\bigr) - \frac{11}{9}
  \right)
  \;.
\end{equation}
The simplification is possible because in the range of medium $Z$
we can apply for the ground state a double expansion over two
parameters:
\begin{eqnarray}
  Z\alpha &\ll& 1\;,\nonumber\\
  \kappa&\gg& 1\;.
  \end{eqnarray}
Here and further we consider only a leading non-relativstic
approximation (i.e., the leading term of the $Z\alpha$ expansion).

Highly excited states in muonic and exotic atoms are of particular
interest for precision measurements because they offer a certain
suppression of the interaction between the nucleus and the
orbiting particle. The $n$ dependence of theoretical expressions,
even of the simplest asymptotics, is not a trivial issue. One can
see from expressions with the generalized hypergeometric function
${_3F_2}$ that while the argument is $\kappa_n^2$, the parameters
are $n$ dependent and in fact in actual situations some are
proportional to $n$.

In particular, the parametrical structure of asymptotic results
for high $\kappa_n$ can be easily understood in the coordinate
representation since the characteristic radius of the potential is
the Compton wave length of an electron $\hbar/m_ec$ and the radius
of atomic states is typically $\hbar n^2/Z\alpha mc$. Thus, the
actual expansion is in $n/\kappa_n$, rather than just in
$1/\kappa_n$. A similar situation is with low $\kappa_n$. Study of
the $n$ dependence and a determination of a real parameter of
expansion are important to find the range of validity of various
asymptotics.

Here we derive a general expression for the Lamb shift at medium
values of the nuclear charge $Z$. Finally the vacuum polarization
correction is presented in the leading nonrelativistic
approximation in the form
\begin{equation}\label{defFH}
  \Delta E(nl) =
  \frac{\alpha}{\pi}
  \,
  (Z\alpha)^2
  \,
  \frac{m c^2}{n^2}
  \,
    F_{nl}(\kappa_n)
  \;.
\end{equation}
The Lamb shift splits the levels with $\Delta n =0$ and $\Delta l
\neq 0$ and for this reason we also consider a specific difference
\begin{equation}\label{phi}
  \Phi_{nll^\prime}(\kappa_n)=F_{nl}(\kappa_n)-F_{nl^\prime}(\kappa_n)
  \;,
\end{equation}
and typically for our calculations $l^\prime=l+1$.

We find in this paper asymptotics for low and high $\kappa_n$ and
determine regions of their validity. We study in more detail
circular states and show that for them the low-$\kappa_n$
expansion is an expansion over $n\cdot\kappa_n = \kappa$, while
the high-$\kappa_n$ asymptotics is actually an expansion over
$n^2/\kappa$.

Additionally to well-defined regions of these expansions
($\kappa\ll1$ or $n^2/\kappa\ll1$) there are also two intermediate
regions:
\begin{itemize}
\item low $\kappa_n$, when $\kappa_n\ll1$, but $n\cdot\kappa_n\sim
1$;
\item high $\kappa_n$, when $\kappa_n\gg1$, but $\kappa_n/n\sim
1$.
\end{itemize}
We discuss behavior of the Uehling correction in these two
specific regions.

\section{The Uehling correction in the nonrelativistic approximation:
general consideration for the Lamb shift} \label{sec:1}

Let us first remind how the Uehling correction is calculated in a
general case. The Lamb shift in muonic atoms is a result of
perturbing the Coulomb potential
\begin{equation}
V_C(r) = -\frac{Z\alpha}{r}
\end{equation}
by the Uehling potential \cite{Schwinger}
\begin{equation} \label{Up}
V_U(r) =
 \frac{\alpha}{\pi}
 \int_0^1
 dv \,
 \frac{v^2(1-v^2/3)}{1-v^2} \,
 \left( -\frac{Z\alpha}{r} e^{-\lambda r} \right)\;,
\end{equation}
where the dispersion `photon' mass
\begin{equation}
  \lambda = \frac{2m_e}{\sqrt{1-v^2}}
\end{equation}
plays a role of the inverse Yukawa radius. Here and for other
calculations in this paper we use relativistic units in which
$\hbar=c=1$, while for final results we restore $c$ and $\hbar$ if
necessary.

The Lamb shift in the nonrelativistic approximation is of the form
\begin{eqnarray} \label{enr}
  \Delta E^{(0)}(nl) &=& \int{dr \, r^2} \vert R_{nl}\vert^2 V_U(r)
  \nonumber\\
  &=& \frac{\alpha}{\pi}\,(Z\alpha)^2 \, \frac{m}{n^2} \, F_{nl}(\kappa_n)
  \;,
\end{eqnarray}
where $R_{nl}(r)$ is the radial part of the Schr\"odinger wave
function in a hydrogen-like atom
\begin{equation}
  \varphi_{nlm}({\bf r})=R_{nl}(r) {\rm Y}_{lm}({\bf r}/r)\;.
\end{equation}
Applying the well-known analytic expression for $R_{nl}(r)$ to
Eq.(\ref{enr}) and integrating over $r$, we obtain (see Eq.~(f.9)
in \cite{III})
\begin{eqnarray}
  &\Delta& E^{(0)} (nl) =
  -\frac{\alpha(Z\alpha)}{2n\,\pi}
  \,
  \left( \frac{2Z\alpha m}{n} \right)^{2l+3}
  \frac{(n+l)!}{(2l+1)!(n-l-1)!}
  \nonumber\\
  &\times&
  \int_0^1 dv \;
  \frac{v^2(1-v^2/3)}{1-v^2}\lambda^{2(n-l-1)}
  \left( \frac{2Z\alpha m}{n} + \lambda \right)^{-2n}
  \\ \nonumber
&\times&
  {_2F_1} \left(
    -n+l+1, -n+l+1;\, 2l+2;\, \left( \frac{2Z\alpha m}{n\lambda} \right)^2
  \right)
  \,.
\end{eqnarray}
After replacing the hypergeometric function by an explicit finite
sum, we integrate over $v$ and arrive at the following expression
for $F_{nl}$:
\begin{eqnarray}\label{eu}
  F_{nl}(\kappa_n) &=&
  -
  \frac{(n+l)!}{(n-l-1)!}
  \sum_{i=0}^{n-l-1}
  \frac{1}{(2l+i+1)!}
  \,
  \frac{1}{i!}
\nonumber\\
  &\times&
  \left( \frac{(n-l-1)!}{(n-l-i-1)!} \right)^2\frac{1}{\kappa_n^{2(n-l-1-i)}}
  \\ \nonumber
  &\times&
  \left[
    K_{1,2(n-l-i),2n}(\kappa_n) - \frac{1}{3} K_{2,2(n-l-i),2n}(\kappa_n)
  \right]
  \,,
\end{eqnarray}
where the integrals
\begin{equation}\label{kabc}
  K_{abc}(\kappa) =
  \int_0^1 dv \, \frac{v^{2a}}{(1-v^2)^{b/2}} \,
  \left( \frac{\kappa\sqrt{1-v^2}}{1+\kappa\sqrt{1-v^2}} \right)^c
  \,.
\end{equation}
can be expressed in general in terms of the generalized
hypergeometric functions (\ref{defKabc}) \cite{CJP98}. Here we
mainly follow our notation in \cite{CJP98}, but the definition of
the integral $K$ (see also \cite{cjp_kg})) is different from the
related integral $I$ there. While in the nonrelativistic limit,
when $\epsilon=0$ and the parameter $c$ is integer,
$K_{a,b,c}(\kappa)=I_{a,b,c}(\kappa)$, in the relativistic case
with non-integer $c$ the notation is
$K_{a,b,c}(\kappa)=I_{a,b,c+2\epsilon}(\kappa,\epsilon)$.

We note that for integer $a, b, c$ the result can be expressed in
terms of elementary functions. Using recursive relations (cf.
\cite{CJP98})
\begin{eqnarray}
  \frac{1}{\kappa^{c+1}}
  K_{a,b,c+1}(\kappa)
  &=&
  -\frac{1}{c}\,
  \frac{\partial}{\partial\kappa}
  \biggl[ \frac{1}{\kappa^c} K_{a,b,c}(\kappa) \biggr]
  \,,\label{rec1}\\
  K_{a,b+1,c+1}(\kappa)
  &=&
  \frac{\kappa^2}{c} \,
  \frac{\partial}{\partial\kappa}
  K_{a,b,c}(\kappa)
  \label{rec2}
\end{eqnarray}
we express the correction for an arbitrary state through the
expression for the ground state
\begin{eqnarray}
  &F_{nl}&(\kappa_n) =
  \frac{(n+l)!}{(n-l-1)!(2n-1)!}
  \sum_{i=0}^{n-l-1}
  \frac{1}{(2l+i+1)!}
  \,
  \frac{1}{i!}
\nonumber\\&\times&
 \left( \frac{(n-l-1)!}{(n-l-i-1)!} \right)^2
  \left(\frac{1}{\kappa_n}\right)^{2(n-l-1-i)}
 \\ \nonumber&\times&
  \left( \kappa_n^2 \frac{\partial}{\partial\kappa_n} \right)^{2(n-l-i-1)}
  \,
  \kappa_n^{2(l+i+1)}
  \,
  \left( \frac{\partial}{\partial\kappa_n} \right)^{2(l+i)}
  \frac{F_{10}(\kappa_n)}{\kappa_n^2}
  \label{fnlf10}
  \,.
\end{eqnarray}

The result for $F_{10}$
\begin{equation}
F_{10}(\kappa)=-K_{122}(\kappa)+\frac13K_{222}(\kappa)\;,
\end{equation}
which follows from Eq.(\ref{eu}), is known in simpler terms and in
particular in terms of elementary functions (see Eq.(\ref{eu1s})).
The general expression (\ref{fnlf10}) now presents a correction
for any states in terms of elementary functions. Such an
expression is also very useful to derive various asymptotics once
we find related asymptotics for $F_{10}(\kappa)$. Another way of
the $F_{nl}$ presentation as a single finite sum can be found in
\cite{soto}.

\section{Asymptotic behavior at large $\kappa_n$}\label{sec:2}

In the case of $\kappa_n\gg1$ we can use asymptotics for the
ground state function $F_{10}$ (cf.~\cite{CJP98} and
Eq.(\ref{as1s}))
\begin{eqnarray}
 F_{10}(\kappa)
 &=&
 - \left[\frac{2}{3}\,\ln\big(2\kappa\big)
  -\frac{11}{9}\right]
 -
 \frac{\pi}{2}\,\frac{1}{\kappa}
 +
 { \frac{3}{2} }
 \frac{1}{\kappa^2}
 \nonumber \\
 &-&
 \frac{2\pi}{3}\,
 \frac{1}{\kappa^3}
 +\left[
 \frac{5}{4}\ln(2\kappa)+
 \frac{1}{16}
 \right]\frac{1}{\kappa^4}
 \\ \nonumber
 &+&
  \left[
 \frac{7}{12}\ln(2\kappa)-\frac{5}{18}
 \right]\frac{1}{\kappa^6}
 + \dots\label{f10long}
\end{eqnarray}

An expression for an arbitrary state can be also derived as an
expansion over $1/\kappa_n$. Here we present few first terms
(cf.~\cite{CJP98,soto})
\begin{eqnarray}
 &F_{nl}&(\kappa_n)
  =
  -\frac{2}{3} \biggl[ \ln \left(2\kappa_n\right) +\psi(1)-\psi(n+l+1) -\frac56 \biggr]
  \nonumber\\
&-&
 \frac{\pi}{2}\,\frac{n}{\kappa_n}
 +\frac12\biggl[{n(2n+1)+(n+l)(n-l-1)}\biggr]\,\frac{1}{\kappa_n^2}
\\ \nonumber
&-&
 \frac{\pi }{9}\biggl[(2n+1)(n+1)+3(n+l)(n-l-1) \biggr]
 \,\frac{n}{\kappa_n^3} + \dots
 \,,
 \label{fnllong}
\end{eqnarray}
%
where $\psi(x)=\Gamma'(x)/\Gamma(x)$ is the logarithmic derivative
of the gamma function and
\[
  \psi(n) - \psi(1) = \sum_{i=1}^{n-1} \frac{1}{i}
\].

The results for the asymptotics of the difference (\ref{phi})
related to the Lamb shift are much simpler than the result for
each level separately:
 \begin{equation}
  \Phi_{n,l-1,l}(\kappa_n)
  =
  -\frac{2}{3}\,\frac{1}{n+l}
  +\frac{l}{\kappa_n^2}
  -\frac{2\pi}{3}\,\frac{n\,l}{\kappa_n^3}
  +\dots
\end{equation}


To test our calculations, we consider a limit $\ln\kappa_n\gg 1$
and find the leading logarithmic term within the effective charge
approach with the help of a substitution
\begin{equation}
  Z\alpha \longrightarrow
  Z\alpha(\kappa_n)
  =
  Z\alpha \left( 1+ \frac{2\alpha}{3\pi} \ln\kappa_n \right)
  \;.
\end{equation}
The result reads
\begin{equation}\label{lamblog}
  F^{\rm log}_{nl}(\kappa_n)= -\frac{2}{3}
  \,  \ln\kappa_n\;.
\end{equation}
The logarithmic contribution vanishes for the Lamb splitting
$\Phi_{n,l-1,l}$. The logarithmic results are in agreement with
the direct calculations above.

\section{Asymptotic behavior at large $\kappa_n$ and large $n$}\label{sec:3}

We note that the asymptotic coefficients depend on $n$ and one may
wonder about their behavior at high $n$.  To study this we apply
the well-known expansion for $\psi(z)$ at high $z$
\[
\psi(z+1)= \ln{z}+\frac{1}{2z} - \frac{1}{12z^2}+\dots
\]

The result for the Uehling correction reads
\begin{eqnarray}
  F_{nl}(\kappa_n)
  &=&-\frac{2}{3} \biggl[ \ln\left(\frac{2\kappa_n}{n+l}\right) -{\cal C} -\frac56
  \nonumber\\
  &-&\frac{1}{2}\frac{1}{(n+l)}+\frac{1}{12}\frac{1}{(n+l)^2}+\dots\biggr]
  \\
  &-&
   \frac{\pi}{2}\,\left(\frac{n}{\kappa_n}\right)
 +\frac{3n^2 -l(l+1)}{2n^2}\,\left(\frac{n}{\kappa_n}\right)^2
  \nonumber\\ \nonumber
&-&
 \frac{\pi}{9}\frac{5n^2 -3l(l+1)-1}{n^2}
 \,\left(\frac{n}{\kappa_n}\right)^3 + \dots
 \,,
 \label{fnllong_a}
\end{eqnarray}
where ${\cal C}= -\psi(1)=0.577\,215\,665\,...$ is Euler's
constant. Certain simplifications are achieved once we do an
assumption on a particular relation between values of $l$ and $n$.

\subsection{Low-$l$ states}\label{subsec:31}

An important feature of the result in Eq.~
(27) is that the parameter of expansion is rather $n/\kappa_n$
than $1/\kappa_n$. For instance, our explicit result for $F_{nl}$
at $n\gg 1$ and low $l$ ($l\ll n$) is
\begin{eqnarray}
 F_{nl}(\kappa_n)&=&-\frac{2}{3}
 \biggl[
 \ln\left(\frac{2\kappa_n}{n}\right)-{\cal C}
 -\frac{5}{6}-\frac{2l+1}{2n}
 \nonumber \\
 &+& \frac{6l(l+1)+1}{12n^2}
 + \dots
 \biggr]-\frac{\pi}{2}\left(\frac{n}{\kappa_n}\right)
 \nonumber\\
 &+&
 \left[\frac{3}{2}-\frac{l(l+1)}{2n^2}\right]\left(\frac{n}{\kappa_n}\right)^2
  \\ \nonumber
 &-&
 \pi\left[\frac{5}{9}-\frac{3l(l+1)-1}{9n^2}\right]\left(\frac{n}{\kappa_n}\right)^3
 +\dots
\end{eqnarray}
We keep here the $l$ dependence in the $1/n^2$ terms in order to
derive a related result for the Lamb splitting
 \begin{eqnarray}\label{lamb_lowl}
  \Phi_{n,l-1,l}(\kappa_n)
  &=&\frac{1}{n}\biggl\{
  -\frac{2}{3}+\frac{2l}{3n}
   \\ \nonumber
   &+&\frac{1}{n}\,\left(\frac{n}{\kappa_n}\right)^2
  - \frac{2\pi}{3n}\,\left(\frac{n}{\kappa_n}\right)^3
  +\dots\biggr\}\;.
\end{eqnarray}
We note that the expansion in (28) and (29) is effectively done in
$n/\kappa_n$. Meanwhile, the leading term in (29) is suppressed by
a factor of $1/n$ and the two first corrections are additionally
suppressed by $1/n$.

\subsection{Near circular states}\label{subsec:32}

After studying $n\gg1$ at low $l$, we turn to another case of
$n\gg1$ at low values of the radial quantum number $n_r=n-l-1\sim
1$. In particular, $n_r=0$ is related to the so-called circular
state. In the limit of high $\kappa_n$ and $n$ we obtain
\begin{eqnarray}\label{fhk}
 &&F_{n,n-n_r-1}(\kappa_n)
  =-\frac{2}{3}\left[\ln\left(\frac{\kappa_n}{n}\right)
 -{\cal C}-\frac{5}{6}+\frac{2n_r+1}{4n}
+ \dots \right]\nonumber\\
 &&-
 \frac{\pi}{2}\,\left(\frac{n}{\kappa_n}\right)
 +
 \left[
 1+\frac{2n_r+1}{2n}+\dots
 \right] \left(\frac{n}{\kappa_n}\right)^2
\\ \nonumber
 &&-\frac{\pi}{9}\left[2+\frac{6n_r+3}{n}+\dots\right]
  \left(\frac{n}{\kappa_n}\right)^3
+{\cal O}\left(\left(\frac{n}{\kappa_n}\right)^4\right)\,.
 \end{eqnarray}
In the same limit the specific difference related to the Lamb
shift is
\begin{eqnarray}\label{phihkn}
  &&\Phi_{n,n-n_r-2,n-n_r-1}(\kappa_n)
  =
  \nonumber\\
  &&\qquad
  =\frac{1}{n}\left\{
  -\frac{1}{3}
  +\left(\frac{n}{\kappa_n}\right)^2
  -2\pi\left(\frac{n}{\kappa_n}\right)^3
  +\dots\right\}\,.
\end{eqnarray}

The difference is suppressed by $1/n$, as well as for low $l$,
but, in contrast to Eq.(\ref{lamb_lowl}), there is no additional
suppression. As a result, we see that the high-$\kappa_n$
expansions above (cf. \cite{CJP98,soto}) are valid only in the
case of $\kappa_n\gg n$, which reduces the range of their
applicability drastically. We consider the case of $\kappa_n\gg1$,
but not $\kappa_n\gg n$ in Sect.~\ref{s:gg_n}.

\section{Asymptotics at low $\kappa$}\label{sec:4}

In principle we are interested in high rather than in low $\kappa$
values, because the problem is related to muonic and exotic atoms.
However, for high $n$, even for $\kappa\gg 1$ we can easily arrive
at a situation when $\kappa_n=\kappa/n\ll 1$ and thus this region
is of interest.

The asymptotic behavior of $F_{nl}(\kappa_n)$ at small values of
$\kappa_n$ was studied in \cite{CJP98} (see also \cite{soto}).
Various approaches can be used for that. One may start from our
expression (\ref{fnlf10}) (cf. \cite{CJP98,soto}), taking into
account that
\begin{eqnarray}
  F_{10}(\kappa) &=&
  -\frac{4\kappa^2}{15}
  +\frac{5\pi\kappa^3}{48}
  -\frac{12\kappa^4}{35}
  +\frac{7\pi\kappa^5}{64}
  -\frac{64\kappa^6}{189}
  \\ \nonumber
  &+&\frac{27\pi\kappa^7}{256}
  -\frac{32\kappa^8}{99}
  +\frac{77\pi\kappa^9}{768}
  -\frac{1536\kappa^{10}}{5005}
  +\dots\;,
\end{eqnarray}
or apply Eq.~(\ref{eu}) with $K_{abc}$ presented in terms of
integral (\ref{kabc}) or of generalized hypergeometric functions
(\ref{defKabc}). Actually, the latter is the most straightforward
way to obtain a low-$\kappa_n$ expansion. In case $\kappa_n\ll1$
the expansionreads
\begin{eqnarray}
 \label{expFsmall}
 &F_{nl}&(\kappa_n)
 =
 -\frac{(n+l)!\;\kappa_n^{2l+2}\,}{(2l+1)!(n-l-1)!}
 \Biggl\{
  \frac{1}{2(l+1)} \, \frac{(2l+4)!!}{(2l+5)!!}
 \nonumber\\
  &-&\pi \, (n\,\kappa_n) \, \frac{1}{2l+3} \, \frac{(2l+5)!!}{(2l+6)!!}
 \\&+&
  (n\,\kappa_n)^2 \left( \frac{4l+5}{l+1}+\frac{l+1}{n^2} \right) \, \frac{1}{4(l+2)} \, \frac{(2l+6)!!}{(2l+7)!!}
\nonumber\\&-&
  \pi (n\,\kappa_n)^3 \left( \frac{4l+7}{l+1}+\frac{3l+5}{n^2} \right) \, \frac{1}{6(2l+5)} \, \frac{(2l+7)!!}{(2l+8)!!}
 \nonumber\\ \nonumber
  &+&{\cal O}((n\kappa_n)^4\bigr)
 \Biggr\}\,.
\end{eqnarray}
The first term of this expansion is obtained in \cite{soto} and is
in agreement with our expression. As one can see, the series is in
fact over $n\cdot\kappa_n=\kappa$ rather than $\kappa_n$. That
sets a condition for applicability of the low-$\kappa_n$
asymptotics as $\kappa_n\ll 1/n$. In particular, it means that the
asymptotics Eq.(\ref{expFsmall}) cannot be applied for Rydberg
states even for the muonic hydrogen, i.e., for the smallest
possible $Z$ ($Z=1$), where $\kappa\sim1.5$ and $\kappa_n\sim
1.5/n$.

\section{High $n$ asymptotic behavior \label{s:gg_n}}\label{sec:5}

We see that while we expand the generalized hypergeometric
function in terms of either $\kappa_n$ or $1/\kappa_n$, the real
parameters of both expansions involve a factor of $n$ directly.
That is due to the increase of the coefficients of the $\kappa_n$-
or $1/\kappa_n$- expansions with $n$ which technically originates
from the expansion of the factor
\begin{equation}\label{nfactor}
\left( \frac{\kappa_n\sqrt{1-v^2}}{1+\kappa_n\sqrt{1-v^2}}
\right)^c
\end{equation}
in the basic integral $K_{abc}(\kappa_n)$, while $c= 2n$.

We note that a consideration of high $n$ is not unrealistic. For
instance, in [neutral] antiprotonic helium for realistic levels
\cite{aHe} we find $Z=2$, $n\simeq 30\gg1$, $\kappa\simeq 27\gg1$,
$\kappa_n \simeq 1$. One of the reasons to study high-$n$ states
is that they very weakly interact with the nucleus, especially if
a value of $l$ is also high. Such an immunity to the
nuclear-structure effects is an advantage from both theoretical
and experimental point of view. Therefore and also because of
simplifications in calculations we consider below circular or near
circular states at $n\gg1$.

\subsection{Limit of low $\kappa_n$ for the near-circular states}\label{subsec:51}

The combination of the $K_{abc}$ integrals which actually enters
the equation for the vacuum-polarization energy shifts is
\begin{equation}
  K_{bc}(\kappa_n)
  =
  K_{1bc}(\kappa_n) - \frac{1}{3} K_{2bc}(\kappa_n)
  \,.
\end{equation}
We find that $b\ll n$ for the near-circular states, and, as long
as we use (\ref{eu}), $c= 2n$ for any state.

Once we know the general expression (\ref{defKabc}) in terms of
$_3F_2$ (cf.~\cite{CJP98}), we can consider in each order of the
$\kappa_n$ expansion only terms leading in $n$ (we did above a
similar procedure to prove for the few first terms of series that
the expansion is over $n\kappa_n$, and not over just $\kappa_n$).

Collecting the leading in $n$ terms we arrive at the result in the
limit $n\gg1$, $\kappa_n\ll 1$ and $b\ll n$
\begin{eqnarray}
 K_{b,2n}(\kappa_n)
 &\simeq&
  \frac{(\kappa_n)^{2n}}{2}
 \;
 B\left( n,\frac32 \right)
 \;
 \Biggl[
 {_2F_1}\left(n, n;\; \frac12 ;\; \kappa_n^2  \right)
 \nonumber\\
 &-&2(n \kappa_n)\;
 {_2F_1}\left(n, n;\; \frac32 ;\; \kappa_n^2  \right)
 \Biggr]
 \,.
\end{eqnarray}
We note that
\begin{eqnarray}
  {}_2F_1(n,n;\; \nu; \;\kappa_n^2) &\simeq& \Gamma(\nu) \; (n \kappa_n)^{1-\nu} \;
  I_{\nu-1}(2n\kappa_n)\;,
  \nonumber\\
  B(n,\nu) &\simeq& \Gamma(\nu) \; n^{-\nu}
\end{eqnarray}
at $\nu\ll n$, where $I_\nu(z)$ is the modified Bessel function.
For the latter one can apply the well-known explicit expressions
for $\nu=1/2, 3/2$ and we arrive at the expression
\begin{equation}\label{KbcLowkn}
 K_{b,2n}(\kappa_n)
 \simeq
 \frac{\sqrt\pi\kappa_n^{2n} }{4n^{3/2}} \, e^{-2n\kappa_n}
  \,.
\end{equation}

To express the correction to energy $F_{nl}(\kappa_n)$ in terms of
the basic integrals $K_{b,2n}(\kappa_n)$ for near-circular states
($n_r=n-l-1\ll n$) we need to transform the related coefficients
in (\ref{eu}) in the limit of high $n$. We note that the integral
$K_{b,2n}(\kappa_n)$ does not depend on $b$ in the leading
$1/n$-approximation and thus the $l$ dependence of the correction
comes from the $l$ dependence of the coefficients of (\ref{eu}).
Eventually we find
\begin{eqnarray}
  F_{nl}(\kappa_n) &\simeq&
  - \frac{\sqrt\pi\,\kappa_n^{2l+2}}{4n^{3/2}}\,
  \frac{(2n)^{n_r}}{n_r!}
  e^{-2n \kappa_n}
  \nonumber\\
  &=&
  - \frac{\sqrt\pi}{4n^{3/2}}\,
  \frac{\kappa_n^{2n}e^{-2n
  \kappa_n}}{n_r!}
  \left(\frac{2n}{\kappa_n}\right)^{n_r}
  \,.
\end{eqnarray}

To conclude this consideration we need to discuss the accuracy and
validity of our derivation. It is valid for $\kappa_n\ll1$ and
$\kappa\sim 1$ and the corrections are of relative order $1/n$. In
the case of $\kappa\ll 1$ it is consistent with the leading term
of the low-$\kappa_n$ expansion (\ref{expFsmall}).

\subsection{The limit of high $\kappa_n$ for the near-circular states}\label{subsec:52}

For $\kappa_n\gg1$ we also consider only near-circular states, for
which $b\sim n_r=n-l-1 \ll n$. We can rewrite $K_{b,2n}(\kappa_n)$
in terms of the basic integrals as follows
\begin{eqnarray}
\nonumber
  &&K_{b,2n}(\kappa_n) =
  \int_0^1 dv \, \frac{v^{2}(1-v^2/3)}{(1-v^2)^{b/2}} \,
  \left( \frac{\kappa_n\sqrt{1-v^2}}{1+\kappa_n\sqrt{1-v^2}} \right)^{2n}
\\
\nonumber
  &&=
  \int_0^1 dv \, \frac{v^{2}(1-v^2/3)}{(1-v^2)^{b/2}} \,
  \exp\left\{2n \ln \left( 1-\frac{1}{1+\kappa_n\sqrt{1-v^2}}
  \right)\right\}
  \,.\label{asy_log}
\end{eqnarray}
If $b\ll n$, we can expand the exponential in the integrand and
find
\begin{equation}
  K_{b,2n}(\kappa_n)
  \simeq
  \int_0^1 dv \, \frac{v^{2}(1-v^2/3)}{(1-v^2)^{b/2}} \,
  e^{-\frac{2n}{\kappa_n\sqrt{1-v^2}}}
  \,,
\end{equation}
that depends upon combination of parameters $\kappa_n/n$ only.
After a substitute of the variable $t=1/\sqrt{1-v^2}$ in this
integral we arrive at the result
\begin{equation}\label{KbcHighkn}
  K_{b,2n}(\kappa_n)
  \simeq
  \int_1^\infty dt \; \frac{\sqrt{t^2-1}}{t^{6-b}} \; \frac{2t^2+1}{3} \; e^{- \frac{2 n
  t}{\kappa_n}}\;.
\end{equation}

Substituting the expression into the sum in Eq.(\ref{eu}), and
neglecting $n-l$ as compared with $n$ in coefficients of the sum,
we obtain
\begin{eqnarray}\label{KbcHighkn2}
  F_{nl}(\kappa_n)
  &\simeq& -\int_1^\infty dt \,
  \frac{\sqrt{t^2-1}}{t^4} \frac{2t^2+1}{3}
  \;e^{- \frac{2 n t}{\kappa_n}}
  \nonumber\\
  &\times&
  \sum_{j=0}^{n_r} \frac{(n_r)!}{(n_r-j)!} \frac{1}{(j!)^2}
 \left(\frac{\sqrt{2n}}{\kappa_n}\,t\right)^{2j}
  \,.
\end{eqnarray}

Similarly to the previous subsection, we find that our derivation
is appropriate for $1/\kappa_n\ll1$ and $n/\kappa_n\sim 1$ and the
result has a relative uncertainty on the order of $1/n$.

\subsection{Comparison of low-$\kappa$ and high-$\kappa$ asymptotics}
\label{comp.sec}

The region of highest interest is the one for high-$\kappa$ is
when $1/\kappa_n\ll1$, but not $n/\kappa_n\ll1$, since for the
opposite situation ($n/\kappa_n\gg1$) we have already known the
proper asymptotic form of the correction. We note that for the
region of interest $\kappa_n/n$ can be about unity or even larger
(e.g., as in the case of $1/\kappa_n\sim {n}^{-1/2}\ll1$ and
$n/\kappa_n\sim {n}^{1/2}\gg1$). In particular, if
$\kappa_n/n\gg1$ the result of the $t$-integration in
Eq.(\ref{KbcHighkn2}) will mainly come from a narrow region
$(t-1)\ll 1$. That means that we can improve the final result once
we consider a complete series for the logarithm in
Section~\ref{subsec:52}, and setting $t=1$ for all terms except of
the leading term of the expansion
\[
  \exp\left\{2n \log \left( 1-\frac{1}{1+\kappa_n\sqrt{1-v^2}}
  \right)\right\}\simeq
\]
\[
 \left( \frac{\kappa_n}{1+\kappa_n} \right)^{2n}\times e^{-
\frac{2 n}{\kappa_n}(t-1)}
  \,.
\]
The final estimation reads for high $\kappa_n$
\begin{equation}\label{KbcHighkn_i}
  K_{b,2n}(\kappa_n)
  \simeq
  \left( \frac{\kappa_n}{1+\kappa_n} \right)^{2n}
  \int_1^\infty dt \; \frac{\sqrt{t^2-1}}{t^{6-b}} \; \frac{2t^2+1}{3} \; e^{- \frac{2 n
  t}{\kappa_n}}
  \,.
\end{equation}

We can also rewrite the result for low $\kappa_n$
(Eq.(\ref{KbcLowkn})) as
\begin{equation}\label{KbcLowkn_i}
 K_{b,2n}(\kappa_n)
 \simeq
 \left( \frac{\kappa_n}{1+\kappa_n} \right)^{2n}\times\frac{\sqrt\pi}{4n^{3/2}}
  \,.
\end{equation}

\begin{figure}
\begin{center}
\resizebox{0.50\textwidth}{!}
{\includegraphics{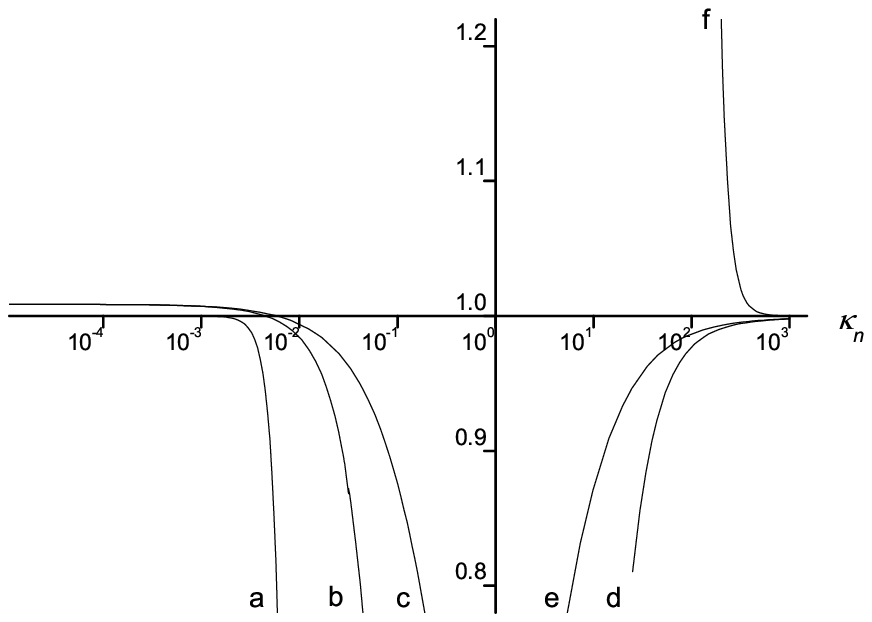}}
\end{center}
\caption{Ratio of different $F_{n,n-1}$ asymptotics and its exact
values for the circular state with $n=100$: (a) -- $n\kappa_n\ll1$
(Eq.(\ref{expFsmall})), (b) -- $\kappa_n\ll1$
(Eq.(\ref{KbcLowkn})), (c) -- $\kappa_n\ll1$
(Eq.(\ref{KbcLowkn_i})), (d) -- $\kappa_n\gg1$
(Eq.(\ref{KbcHighkn})), (e) -- $\kappa_n\gg1$
(Eq.(\ref{KbcHighkn_i})), (f) -- $\kappa_n/n\gg1$
(Eq.(\ref{fhk})); the horizontal axis is related to $F_{n,n-1}$. }
\label{asym100.fig}
\end{figure}

Comparing those two asymptotics for the circular-state correction,
we find that the integral $K_{2,2n}(\kappa_n)$ can be presented as
a product of a factor
\begin{equation}\label{thefactor}
 \left( \frac{\kappa_n}{1+\kappa_n} \right)^{2n}
\end{equation}
and a smooth function. The factor is varying in an extremely broad
region of $\kappa_n$: from low $\kappa_n$ ($\kappa_n\ll1/n$) to
large $\kappa_n$ ($\kappa_n\gg n$), while the smooth function
changes from being proportional to $n^{-3/2}$ to
$\ln({\kappa}/n^2)$. Because of so smooth behavior we expect that
the asymptotics with the explicit factor Eq.(\ref{thefactor}) can
be successfully applied for a somewhat larger region, however,
their accuracy there is unclear. Various asymptotics are compared
to the exact result for $n=100$ in Fig.~\ref{asym100.fig}. In
particular we see that an explicit presentation of the factor of
$(\kappa_n/(1+\kappa_n))^{2n}$ really improves agreement between
the asymptotics and the exact solution.

\section{Other states}\label{sec:6}

Above we obtained the high-$n$ asymptotic expressions in two
specific regions of parameters where $\kappa_n\sim1/n$ and
$\kappa_n\sim n$ for the circular and the near-circular states
only. We are also interested in finding asymptotics in these
regions that are valid for low $l$.

For low $\kappa_n$ and $l\ll n$ we can use an approximate relation
\begin{eqnarray}
  &&K_{2(n-l-i),2n}(\kappa_n) \simeq
  \\ \nonumber
  &&\kappa_n^{2n}\,\int_0^1 dv \, v^{2}
  \left(1-\frac{v^2}{3}\right)\left(1-v^2\right)^{l+i}
  \exp\left\{-2n \kappa_n\sqrt{1-v^2}\right\}
  \;,
\end{eqnarray}
and, neglecting $l$ as compared with $n$ in coefficients of the
sum Eq.(\ref{eu}), obtain an approximation
\begin{eqnarray}
&&F_{nl}(\kappa_n)\simeq -\frac{\left( {n}{\kappa_n}
\right)^{2(l+1)}}{n}\int_0^1 dv \, v^{2}
  \left(1-\frac{v^2}{3}\right)\left(1-v^2\right)^{l}
  \nonumber\\
  &&\times
  \exp\left\{-2n \kappa_n\sqrt{1-v^2}\right\}
  \sum_{i=0}^{n-l-1}\frac{\left(n\kappa_n\sqrt{1-v^2}\right)^{2i}}{i!\,(2l+i+1)!}\,
  \;.
\end{eqnarray}
We can see that the asymptotic depends upon a combination of
parameters $n\kappa_n$, confirming the above-mentioned fact that
it is the real parameter of expansion at low $\kappa_n$.

In the other region corresponding to $\kappa_n \sim n$ we do not
see a simple way to find a proper asymptotic form for low-$l$
states.

\section{Summary}\label{sec:7}

Concluding, we have to briefly discuss corrections to the results
derived. Since the parameter $m/M$ (here $M$ is the mass of the
nucleus) in muonic, pionic and other exotic atoms is not as small
as in conventional atoms, an important question is the accuracy of
our results obtained in the external field approximation, i.e. in
the limit $m/M=0$. In antiprotonic atoms such effects are even
more important than in the muonic case. The higher $m/M$
corrections, which are quite important for muonic atoms, can be
easily taken into account for the Lamb shift in the
nonrelativistic approximation by substituting the mass of the
orbiting particle $m$ for the reduced mass $m_R=m M/(m+M)$ in
Eq.(\ref{defFH}) and $\kappa$ for $\kappa_R=Z\alpha m_R/m_e$.

\begin{table}
\def\arraystretch{1.5}
 \begin{center}
 \caption{Asymptotics at $x\gg1$ for the Uehling correction for the
lowest $s$ states. The correction is presented in terms of a
dimensionless function $F_{nl}$: $\Delta E(nl) =
({\alpha}/{\pi})\,((Z\alpha)^2{m c^2}/{n^2}) \times
F_{nl}(\kappa_n)$} \label{tabNR}
 \begin{tabular}{|c||l|}
 \hline $n$ &
  \hfil $F_{n0}(x)$ \hfil
 \\
 \hline
 1 & $-\frac{2}{3}\ln(2x) + \frac{11}{9} - \frac{\pi}{2x} +
 \frac{3}{2x^2}-\frac{2\pi}{3x^3}+\ldots$
 \\
 \hline
 2 & $-\frac{2}{3}\ln(2x) + \frac{14}{9} - \frac{\pi}{x} +
 \frac{6}{x^2}-\frac{14\pi}{3x^3}+\ldots$  \\
 \hline
 3 & $-\frac{2}{3}\ln(2x) + \frac{16}{9} - \frac{3\pi}{2x} +
 \frac{27}{2x^2}-\frac{46\pi}{3x^3}+\ldots$ \\
 \hline
 4 & $-\frac{2}{3}\ln(2x) + \frac{35}{18} - \frac{2\pi}{x} +
 \frac{24}{x^2}-\frac{36\pi}{x^3}+\ldots$ \\
 \hline
 5 & $-\frac{2}{3}\ln(2x) + \frac{187}{90} - \frac{5\pi}{2x} +
 \frac{75}{2x^2}-\frac{70\pi}{x^3}+\ldots$
 \\
 \hline
 \end{tabular}
 \end{center}
 \end{table}

The functions $F_{nl}$  are presented above for arbitrary $nl$ in
a closed analytic form in various ways. Certain asymptotics are
also presented. The results at $\kappa_n\gg 1$ for the lowest
states are summarized in Table~\ref{tabNR}. They are simple and
transparent. The Lamb shift result in Eq.(\ref{eu}) is obtained in
the nonrelativistic approximation and is valid for any
hydrogen-like atom as far as the relativistic corrections can be
neglected. Some asymptotic results for the splitting of levels
with $\Delta l=1$ at $\kappa_n\gg n$ for some low lying states are
summarized in Table~\ref{tabDIFF}.

\begin{table}
\def\arraystretch{1.5}
 \begin{center}
 \caption{Asymptotics of the Lamb-shift-induced difference at
$x\gg1$ for the lowest states presented in terms of a
dimensionless function $\Phi_{nab}$: $\Delta E(n,l)- \Delta
E(n,l^\prime) = ({\alpha}/{\pi})\,((Z\alpha)^2{m c^2}/{n^2})
\times \Phi_{nll^\prime}(\kappa_n)$} \label{tabDIFF}
 \begin{tabular}{|c||l|l|l|l|}
 \hline $n$ &
  \hfil $\Phi_{n01}(x)$ \hfil &
  \hfil $\Phi_{n12}(x)$ \hfil &
  \hfil $\Phi_{n23}(x)$ \hfil &
  \hfil $\Phi_{n34}(x)$ \hfil
 \\
 \hline
 2 & $-\frac{2}{9} + \frac{1}{x^2}$ &&&
 \\& $ -\frac{4\pi}{3x^3}+\ldots$
 &&&\\
 \hline
 3 & $-\frac{1}{6} + \frac{1}{x^2}$
 & $-\frac{2}{15} + \frac{2}{x^2}$
 &&\\
 & $-\frac{2\pi}{x^3}+\ldots$ & $-\frac{4\pi}{x^3}+\ldots$
  &&\\
 \hline
 4  & $-\frac{2}{15} + \frac{1}{x^2}$
 & $-\frac{1}{9} + \frac{2}{x^2}$
 & $-\frac{2}{21} + \frac{3}{x^2}$
 &\\
 &$-\frac{8\pi}{3x^3}+\ldots$
 &$-\frac{16\pi}{3x^3}+\ldots$
 &$-\frac{8\pi}{x^3}+\ldots$
 &\\
 \hline
 5
 & $-\frac{1}{9} + \frac{1}{x^2}$
 & $-\frac{2}{21} + \frac{2}{x^2}$
 & $-\frac{1}{12} + \frac{3}{x^2}$
 & $-\frac{2}{27} + \frac{4}{x^2}$
 \\
 &$-\frac{10\pi}{3x^3}+\ldots$
 &$-\frac{20\pi}{3x^3}+\ldots$
 &$-\frac{10\pi}{x^3}+\ldots$
 &$-\frac{40\pi}{3x^3}+\ldots$
 \\
 \hline
\end{tabular}
\end{center}
\end{table}

\begin{figure}
\begin{center}
\resizebox{0.50\textwidth}{!}
{\includegraphics{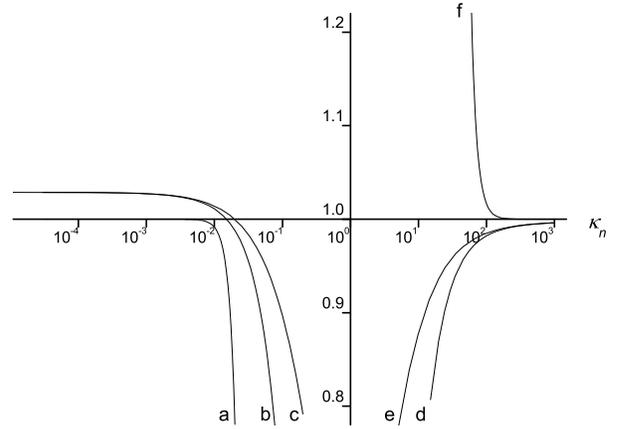}}
\end{center}
\caption{Ratio of different $F_{n,n-1}$ asymptotics and its exact
values for the circular state with $n=30$: (a) -- $n\kappa_n\ll1$
(Eq.(\ref{expFsmall})), (b) -- $\kappa_n\ll1$
(Eq.(\ref{KbcLowkn})), (c) -- $\kappa_n\ll1$
(Eq.(\ref{KbcLowkn_i})), (d) -- $\kappa_n\gg1$
(Eq.(\ref{KbcHighkn})), (e) -- $\kappa_n\gg1$
(Eq.(\ref{KbcHighkn_i})), (f) -- $\kappa_n/n\gg1$
(Eq.(\ref{fhk})); the horizontal axis is related to $F_{n,n-1}$.}
\label{asym30.fig}
\end{figure}

We studied the applicability of naive low-$\kappa$ and
high-$\kappa$ asymptotics and found that the region where they are
valid strongly depends on $n$. For high $n$ we considered some
additional asymptotics (see., e.g., Fig.~\ref{asym30.fig} where
the results are presented for a realistic value of $n=30$). We
found a sum of the leading terms for $n\gg1$ expansions for both
low-$\kappa$ and high-$\kappa$ cases. In particular, we found that
most of the change by orders of magnitude of the Uehling
correction in the circular states can be presented in terms of a
simple factor $ \Bigl( {\kappa_n}/({1+\kappa_n}) \Bigr)^{2n}$
which is multiplied by a smooth function.

All results are obtained in the leading nonrelativistic approach
and corrections due to that are of relative order $(Z\alpha)^2$.
The relativistic effects will be considered elsewhere.

\section*{Acknowledgements}\label{sec:acknowl}

This work was supported in part by RFBR (grants \#\# 03-02-16843
and 03-02-04029) and DFG (grant GZ 436 RUS 113/769/0-1). Part of
the work was performed during visits of VGI and EYK at the
Max-Planck-Institut f\"ur Quantenoptik and we are grateful for
their hospitality.

\end{document}